# Likelihood confidence intervals for misspecified Cox models

Yongwu Shao[1], Xu Guo[1]

12 August 2025

## Abstract

The robust Wald confidence interval (CI) for the Cox model is commonly used when the model may be misspecified or when weights are applied.  However it can perform poorly when there are few events in one or both treatment groups, as may occur when the event of interest is rare or when the experimental arm is highly efficacious.  For instance, if we artificially remove events (assuming more events are unfavorable) from the experimental group, the resulting upper CI may increase.  This is clearly counter-intuitive as a small number of events in the experimental arm represents stronger evidence for efficacy.

It is well known that, when the sample size is small to moderate, likelihood CIs are better than Wald CIs in terms of actual coverage probabilities closely matching nominal levels.  However, a robust version of the likelihood CI for the Cox model remains an open problem.  For example, in the SAS procedure PHREG, the likelihood CI provided in the outputs is still the regular version, even when the robust option is specified.  This is obviously undesirable as a user may mistakenly assume that the CI is the robust version.

In this article we demonstrate that the likelihood ratio test statistic of the Cox model converges to a weighted chi-square distribution when the model is misspecified.  The robust likelihood CI is then obtained by inverting the robust likelihood ratio test.  The proposed CIs are evaluated through simulation studies and illustrated using real data from an HIV prevention trial.

## 1   Introduction

The robust (sandwich, Huber[1] [1967]) variance matrix for the Cox proportional hazard model[2] was first proposed by Lin and Wei[3] (1989) to account for potential model misspecification.  The robust Wald test and the robust Wald confidence intervals (CIs) can then be constructed using the robust standard errors.  Since their introduction, these methods have been widely adopted due to their reliability and robustness.  In recent years, with the increasing use of propensity score weights[4], these methods have become even more popular, as weighted observations necessitates the use of robust standard errors.

---
[1] Gilead Sciences, Foster City, California, USA
**Correspondence**
Yongwu Shao, Gilead Sciences, 333 Lakeside Dr, Foster City, CA 94404, USA, Email: ywshao@gmail.com

One issue of the robust Wald CI is that it may perform poorly when there are few events in one or both treatment groups, as may be the case when the event of interest is rare or when the experimental arm is highly efficacious.  Let us take the HPTN 084 trial as an example.  It is a phase 3, randomized, double-blind, double-dummy, superiority trial evaluating the safety and efficacy of injectable cabotegravir compared with daily oral tenofovir diphosphate plus emtricitabine (TDF-FTC) for HIV infection in HIV-uninfected women (Delany-Moretlwe et al[5], 2022). A total of 3,224 eligible participants were randomly assigned (1:1) to either active cabotegravir with TDF-FTC placebo (cabotegravir group) or active TDF-FTC with cabotegravir placebo (TDF-FTC group). The primary efficacy endpoint was incident HIV infection, and the primary efficacy analysis was based on a proportional hazards model with treatment group as a covariate (to be analyzed using the Wald method). At the end of the trial, 40 incident infections were observed over 3,898 person-years, 4 in the cabotegravir group and 36 in the TDF-FTC group.  Based on the Kaplan-Meier curves of the two treatments (Figure 3 in Delany-Moretlwe et al[5], 2022), there might be a potential violation of the proportional hazard assumption, therefore it makes more sense to use a robust standard error instead to account for possible model mis-specifications.  The estimated hazard ratio using the robust Wald CI for cabotegravir vs TDF-FTC is 0.12 (95% CI: 0.05-0.31), which is significantly less than 1 ($p<0.0001$).

*Figure 1. Hypothetical upper (100-0.033)% CI based on the robust Wald CI for the HPTN 084 trial, with the number of HIV infections in the Cabotegravir arm varying between 1 and 10 instead, while keeping the number of events in the TDF-FTC arm as observed.*

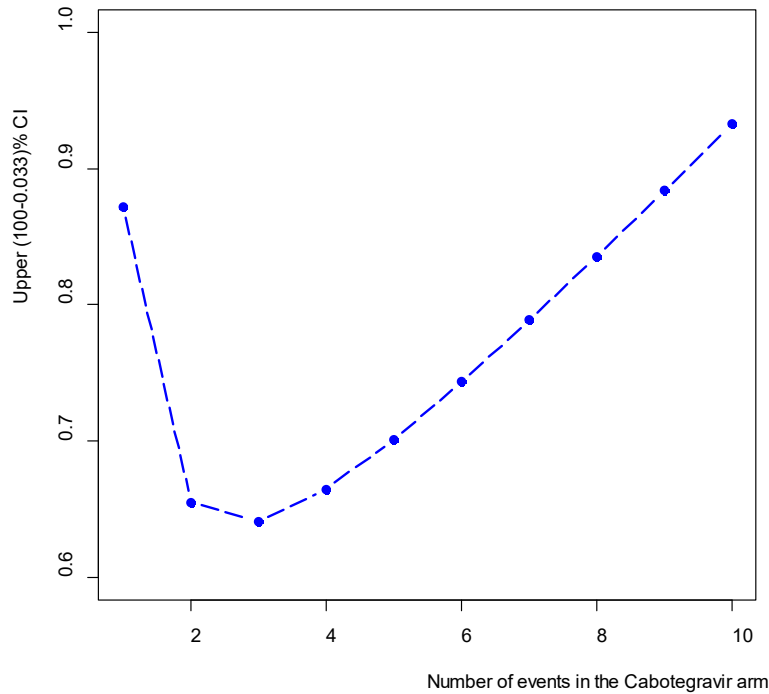

The robustness of the analysis has been assessed in a sensitivity analysis where the number of infections in the cabotegravir group varies between 0 and 10. To reduce the number of infections, we remove the 4 observed events in the same order in which they occurred relative to the baseline visit. To increase the number of events, we add one event to each scheduled cabotegravir injection visit in ascending order (ie, Week 9, 17, 25, etc), until we reach the desired number of events. (The choice of timepoints in adding or removing events have minimal impact on the results.) Since the interim alpha at the time of analysis is 0.00066 two-sided, we focus on the upper (100-0.033)% CI. Figure 1 shows the resulting upper Wald CI as a function of the number of infections in the cabotegravir group. Surprisingly, the function is not monotone increasing – the upper Wald CI first decreases and then increases, with a minimum at 3. A similar issue is noted for the one-sided p-value based on the robust Wald test. The results of this sensitivity analysis are clearly counter-intuitive as a smaller number of infections in the cabotegravir group represents stronger evidence for the efficacy of cabotegravir.

One possible solution to the above problem is to use the likelihood CI (by inverting the likelihood ratio test), as proposed by Venzon and Moolgavkar[6] (1988) for the Cox model. It is well known that, when the sample size is small to moderate, likelihood CIs are better than Wald CIs in terms of actual coverage probabilities coming close to matching nominal levels[7]. Recent evidence indicates that likelihood CIs perform much better than Wald CIs in small and censored samples[8]. Likelihood CIs has been suggested as the method of choice due to its superior performance in small samples typically encountered in applications[8,9].

However, even though the robust CI and the regular CI can differ significantly, especially when weights are applied, a robust version of the likelihood CI for the Cox model has remained an open problem. For instance, in the SAS procedure PHREG, the likelihood CI provided in the outputs is still the regular version proposed by Venzon and Moolgavkar[6], even when the robust option is specified. This is obviously undesirable as a user may mistakenly assume that the CI is the robust version, which can differ substantially in practice.

It should be noted that, for general parametric models, the asymptotic distribution for the likelihood ratio test statistic has been independently derived by Foutz and Srivastava[10] (1977) and Kent[11] (1982) under model violations. But as Lin and Wei (1989) pointed out, the Cox model is a semi-parametric model, and the results regarding the robust test statistics in parametric settings do not directly apply here, as the asymptotic distribution is more complicated than their parametric counterpart.

In this manuscript, we demonstrate that when the Cox model is misspecified, the usual likelihood ratio test statistic converges to a weighted chi-squared distribution, where the weights can be consistently estimated. When the parameter of interest is one-dimensional, the asymptotic distribution simplifies to a scaled chi-squared distribution. The robust likelihood CI is then obtained by inverting the robust likelihood ratio test. By construction, the p-value for robust likelihood ratio test is monotone increasing as a function of the number of events in the experimental arm, as is the upper limit of the robust likelihood CI for the hazard ratio. Simulation results indicates that the robust likelihood CI provides adequate coverage.

## 2   Asymptotic results for the Wald test by Lin and Wei (1989)

Since our results are built based on Lin and Wei (1989), in this section we give a brief review of their main results, as well as the notations that they have used.

Let $Z(t) = \left(Z_1(t), \ldots, Z_p(t)\right)^T$ be a $p$ vector of covariates which could be time-dependent. The Cox proportional hazards model posits that the hazard function $\lambda(t)$ can be written as:
$$\lambda(t; Z) = \lambda_0(t) \exp\{\beta^T Z(t)\}, \tag{1}$$
where $\beta = (\beta_1, \ldots, \beta_p)^T$ represents a $p$-dimensional vector of unknown regression coefficients, and $\lambda_0(t)$ denotes an unspecified baseline hazard function. Let $X_1, \ldots, X_n$ represent $n$ possibly right-censored failure times, with $(Z_1, \ldots, Z_n)$ as the corresponding covariance vectors, where $Z_i$ is observed over the interval $[0, X_i]$. The censoring is assumed to be non-informative. The maximum partial likelihood estimator (MPLE), $\hat{\beta}$, is defined as the value that maximizes the partial log likelihood function (Cox, 1972[2])
$$l(\beta) = \sum_{i=1}^{n} \delta_i \left( \beta^T Z_i(X_i) - \log\left[\sum_{j \in R_i} \exp\{\beta^T Z_j(X_j)\}\right] \right), \tag{2}$$

where $R_i$ denotes the set of labels assigned to the individuals at risk at time $X_i^-$, with $\delta_i$ taking the value of one if $X_i$ represents an observed failure time and zero otherwise.

For the $i$th individual, let $\lambda_i(t)$ be the true hazard function, and $Y_i(t) = I\{X_i \geq t\}$, where $I\{\cdot\}$ denotes the indicator function. Assume that $(X_i, \delta_i, Z_i)(i = 1, \ldots, n)$ are $n$ independent and identically distributed (iid) realizations of $(X, \delta, Z)$. Define the following notations:

$$S^{(r)}(t) = n^{-1} \sum_{i=1}^{n} Y_i(t)\lambda_i(t)Z_i(t)^{\otimes r}$$

$$s^{(r)}(t) = E\{S^{(r)}(t)\}$$

$$S^{(r)}(\beta, t) = n^{-1} \sum_{i=1}^{n} Y_i(t)\exp\{\beta^T Z_i(t)\}Z_i(t)^{\otimes r}$$

$$s^{(r)}(\beta, t) = E\{S^{(r)}(\beta, t)\}$$

for $r = 0,1,2$, where for a column vector $a$, $a^{\otimes 2}$ denotes the matrix $aa'$, $a^{\otimes 1}$ represents the vector $a$, $a^{\otimes 0}$ corresponds to the scalar 1 and the expectations are taken based on the true model of $(X_i, \delta_i, Z_i)(i = 1, \ldots, n)$.

Let (1) be the working model for $(X_i, \delta_i, Z_i)(i = 1, \ldots, n)$. The logarithm of the partial likelihood function can be expressed as $l(\beta) = \sum_{i=1}^{n} \delta_i [\beta^T Z_i(X_i) - \log\{S^{(0)}(\beta, X_i)\}]$. The corresponding score function is $U(\beta) = \sum_{i=1}^{n} \delta_i [Z_i(t) - \frac{S^{(1)}(\beta, X_i)}{S^{(0)}(\beta, X_i)}]$. Then $\hat{\beta}$ converges in probability to a $p$ vector of constants $\beta^*$, which is the unique solution to the system of equations

$$\int_0^\infty s^{(1)}(t)dt - \int_0^\infty \frac{s^{(1)}(\beta, t)}{s^{(0)}(\beta, t)} s^{(0)}(t)dt = 0, \qquad (3)$$

provided that $A(\beta^*)$ is positive definite, where $A(\beta) = \int_0^\infty \left\{\frac{s^{(2)}(\beta,t)}{s^{(0)}(\beta,t)} - \frac{s^{(1)}(\beta,t)^{\otimes 2}}{s^{(0)}(\beta,t)^2}\right\} s^{(0)}(t)dt$.

Let $N_i(t) = I\{X_i \leq t, \delta_i = 1\}$, and $\bar{N}(t) = \sum N_i(t)$, $\tilde{F}_n(t) = \bar{N}(t)/n$, and $\tilde{F}(t) = E\{\tilde{F}_n(t)\}$. Let

$$w_i(\beta) = \int_0^\infty \left\{Z_i(t) - \frac{s^{(1)}(\beta,t)}{s^{(0)}(\beta,t)}\right\} dN_i(t) - \int_0^\infty \frac{Y_i(t)\exp\{\beta^T Z_i(t)\}}{s^{(0)}(\beta,t)} \left\{Z_i(t) - \frac{s^{(1)}(\beta,t)}{s^{(0)}(\beta,t)}\right\} d\tilde{F}(t).$$

Lin and Wei (1989) showed that

$$n^{1/2} A(\beta^*)(\hat{\beta} - \beta^*) = \sum_{i=1}^{n} w_i(\beta^*) + o_p(1). \qquad (4)$$

Notice that $w_i(\beta)$ $(i = 1, \ldots, n)$ are iid. Let $B(\beta^*) = E\{w_i(\beta^*)^{\otimes 2}\}$. Therefore $n^{1/2}(\hat{\beta} - \beta^*)$ is asymptotically normal with zero mean and covariance matrix

$$V(\beta^*) = A^{-1}(\beta^*)B(\beta^*)A^{-1}(\beta^*). \qquad (5)$$

Now, let

$$W_i(\beta) = \delta_i \left\{ Z_i(X_i) - \frac{S^{(1)}(\beta, X_i)}{S^{(0)}(\beta, X_i)} \right\} - \sum_{j=1}^{n} \frac{\delta_j Y_i(X_j) \exp\{\beta^T Z_i(X_j)\}}{n S^{(0)}(\beta, X_j)} \times \left\{ Z_i(X_j) - \frac{S^{(1)}(\beta, X_j)}{S^{(0)}(\beta, X_j)} \right\}.$$

Let

$$\hat{A}(\beta) = n^{-1} \sum_{i=1}^{n} \delta_i \left\{ \frac{S^{(2)}(\beta, X_i)}{S^{(0)}(\beta, X_i)} - \frac{S^{(1)}(\beta, X_i)^{\otimes 2}}{S^{(0)}(\beta, X_i)} \right\}$$

be a consistent estimator of $A(\beta)$. In addition, let $\hat{B}(\beta) = n^{-1} \sum W_i(\beta)^{\otimes 2}$, and let $\hat{V}(\beta) = \hat{A}^{-1}(\beta) \hat{B}(\beta) \hat{A}^{-1}(\beta)$. Then Lin and Wei (1989) obtained the following theorem.

*Theorem 1 (Lin and Wei).* The random vector $n^{1/2}(\hat{\beta} - \beta^*)$ is asymptotically normal with mean 0 and with a covariance matrix $V(\beta)$ defined by (5), which can be consistently estimated by $\hat{V}(\beta)$.

## 3   Asymptotical distribution of the likelihood ratio test statistic

By the application of the Taylor's Theorem on the log likelihood function at the point $\hat{\beta}$, we have

$$l(\beta^*) = l(\hat{\beta}) + U(\hat{\beta})(\beta^* - \hat{\beta}) - \frac{n}{2}(\beta^* - \hat{\beta})^T \hat{A}(\tilde{\beta})(\beta^* - \hat{\beta})$$

where $\tilde{\beta}$ is on the line segment between $\beta^*$ and $\hat{\beta}$. The consistency of $\hat{A}(\tilde{\beta})$ and $A(\beta^*)$ has been established by Lin and Wei (1989) using techniques of Andersen and Gill[12] (1982). Since $\hat{\beta}$ is the MPLE, we have $U(\hat{\beta}) = 0$, and

$$l(\hat{\beta}) = l(\beta^*) + \frac{n}{2}(\hat{\beta} - \beta^*)^T A(\beta^*)(\hat{\beta} - \beta^*) + o_p(1). \tag{6}$$

Plug in (4) and we have

$$l(\hat{\beta}) = l(\beta^*) + \frac{1}{2n} \left\{ \sum_{i=1}^{n} w_i(\beta) \right\}^T A^{-1}(\beta^*) \left\{ \sum_{i=1}^{n} w_i(\beta) \right\} + o_p(1). \tag{7}$$

Now we partition $\beta$ into $\varphi$ and $\eta$, ie, $\beta = \begin{pmatrix} \varphi \\ \eta \end{pmatrix}$, where $\varphi$ is our parameter of interest with $\dim(\varphi) = k$, and $\eta$ is a nuisance parameter with $\dim(\eta) = p - k$. Let $\begin{pmatrix} \varphi^* \\ \eta^* \end{pmatrix} = \beta^*$. Let $\hat{\eta}_*$ be the constrained MPLE of $\eta$ under $\varphi = \varphi^*$. Then following a similar argument of (6), we have

$$l\begin{pmatrix} \varphi^* \\ \hat{\eta}_* \end{pmatrix} = l(\beta^*) + \frac{n}{2}(\hat{\eta}_* - \eta^*)^T A_{22}(\beta^*)(\hat{\eta}_* - \eta^*) + o_p(1), \tag{8}$$

where $A_{22}(\beta^*)$ is the lower bottom $(k - p) \times (k - p)$ sub-matrix of $A(\beta^*)$. By Theorem 1 we have

$$n^{1/2} A_{22}(\beta^*)(\hat{\eta}_* - \eta^*) = \sum_{i=1}^{n} u_i(\beta^*) + o_p(1), \tag{9}$$

where $u_i(\beta^*)$ is the lower $p - k$ elements of the vector $w_i(\beta^*)$. Plug (9) into (8), we have

$$l\begin{pmatrix}\varphi^*\\ \hat{\eta}_*\end{pmatrix} = l(\beta^*) + \frac{1}{2n}\left\{\sum_{i=1}^n w_i(\beta)\right\}^T \begin{bmatrix} 0_{k\times k} & 0 \\ 0 & \{A_{22}(\beta^*)\}^{-1}\end{bmatrix}\left\{\sum_{i=1}^n w_i(\beta)\right\} + o_p(1).$$

Let $C = A^{-1}(\beta^*)$, and partition $C$ into blocks $C = \begin{bmatrix} C_{11} & C_{12} \\ C_{21} & C_{22}\end{bmatrix}$, then $\{A_{22}(\beta^*)\}^{-1} = C_{22} - C_{21}C_{11}^{-1}C_{12}$, hence we have

$$A^{-1}(\beta^*) - \begin{bmatrix} 0_{k\times k} & 0 \\ 0 & \{A_{22}(\beta^*)\}^{-1}\end{bmatrix} = \begin{bmatrix} C_{11} & C_{12} \\ C_{21} & C_{21}C_{11}^{-1}C_{12}\end{bmatrix} = \begin{pmatrix}C_{11}\\C_{21}\end{pmatrix}C_{11}^{-1}(C_{11} \ \ C_{12})$$

Let $2\Delta = 2\left\{l(\hat{\beta}) - l\begin{pmatrix}\varphi^*\\\hat{\eta}_*\end{pmatrix}\right\}$ be the likelihood ratio test statistic. We have

$$2\Delta = n^{-1}\left\{\sum_{i=1}^n w_i(\beta)\right\}^T \begin{pmatrix}C_{11}\\C_{21}\end{pmatrix}C_{11}^{-1}(C_{11} \ \ C_{12})\left\{\sum_{i=1}^n w_i(\beta)\right\} + o_p(1)$$

$$= n^{-1}\text{tr}\left[\begin{pmatrix}C_{11}\\C_{21}\end{pmatrix}C_{11}^{-1}(C_{11} \ \ C_{12})\left\{\sum_{i=1}^n w_i(\beta)\right\}^{\otimes 2}\right] + o_p(1)$$

Since $n^{-1/2}\sum_{i=1}^n w_i(\beta)$ converges to a normal distribution with zero mean and variance $B(\beta^*)$, the likelihood ratio test statistic $2\Delta$ converges to a weighted summation of $k$ chi-square distributions, with the weights being the non-zero eigenvalues of $\begin{pmatrix}C_{11}\\C_{21}\end{pmatrix}C_{11}^{-1}(C_{11} \ \ C_{12})B(\beta^*)$, which are the same as the non-zero eigenvalues of $C_{11}^{-1}(C_{11} \ \ C_{12})B(\beta^*)\begin{pmatrix}C_{11}\\C_{21}\end{pmatrix}$. Note that $(C_{11} \ \ C_{12})B(\beta^*)\begin{pmatrix}C_{11}\\C_{21}\end{pmatrix}$ is simply the top-left $k \times k$ submatrix of $V(\beta^*)$, which we denote by $V_{11}(\beta^*)$. Hence we get the following theorem.

*Theorem 2.* The likelihood ratio test statistic $2\Delta$ is asymptotically distributed as

$$\sum_{j=1}^k \delta_j \chi_j^2(1),$$

where $\chi_j^2(1)$ are independent chi-square distributions with one degree of freedom, $\delta_j$'s are the eigenvalues of the matrix $V_{11}(\beta^*)[\{A^{-1}(\beta^*)\}_{11}]^{-1}$, which can be consistently estimated by $\hat{V}_{11}(\beta)\left[\{\hat{A}^{-1}(\beta)\}_{11}\right]^{-1}$. Here $[\cdot]_{11}$ denotes the top-left $k \times k$ sub-matrix of the corresponding matrix.

## 4 Likelihood confidence intervals

In practice people are often interested in a single parameter like the treatment effect, ie, $\dim(\varphi) = 1$. In this case the matrix $\hat{V}_{11}(\beta)\left[\{\hat{A}^{-1}(\beta)\}_{11}\right]^{-1}$ reduces to a scale and the likelihood ratio test statistic converges to a scaled chi-square distribution.

The likelihood CI can be constructed by inverting the likelihood ratio test in Theorem 2 by defining the CI as the region where 2Δ is less than a critical number, similar to what was done in Venzon and Moolgavkar (1988). This CI could have unequal tail probabilities. In clinical trials, people are often more interested in CIs with equal tail probabilities, as this allows for one-sided statistical significance statements, such as a positive treatment effect. Therefore we will focus on this scenario instead. To do this we need to develop the one-sided test for the hypothesis $H_0: \varphi = \varphi_0$ versus $H_a: \varphi < \varphi_0$ for a given $\varphi_0$.

Cox and Hinkley[13] (1979) introduced the one-sided log-likelihood ratio test statistic $\text{sign}(\hat{\varphi} - \varphi_0)\sqrt{2\Delta}$, where $\hat{\varphi}$ is the unconstrained MPLE for $\varphi$, and $\text{sign}()$ is the usual sign function which takes value of 0, 1, -1, respectively, when the argument is 0, positive, negative. They showed that under the null hypothesis, the above test statistic converges to a standard normal distribution, assuming the model is correctly specified.

When the model assumption is violated, 2Δ converges to a scaled chi-square distribution instead, hence we modify the test statistic to

$$\text{sign}(\hat{\varphi} - \varphi_0)\sqrt{2\Delta\{\hat{V}_{11}(\beta)\}^{-1}\{\hat{A}^{-1}(\beta)\}_{11}}$$

which converges to a standard normal distribution by Theorem 2. This is our proposed one-sided likelihood ratio test statistic for misspecified Cox model.

Having developed the one-sided test, we can use the duality of hypothesis testing and interval estimation (Rohatgi[14], 2013, pp. 224-225) to derive the confidence interval. Specifically, the p-value obtained from the above procedure can be seen as a continuous decreasing function of $\varphi_0$. And as $\varphi_0$ varies from zero to infinity, the p-value will decrease from one to zero. To get the lower $1 - \alpha$ CI, we simply find an $\hat{\varphi}_L$ such that the p-value is exactly $\alpha$. In R this can be implemented using the built-in root-finding function uniroot(). The upper $1 - \alpha$ CI can be computed similarly.

# 5 Simulations

Extensive empirical studies have been conducted to evaluate the properties of the robust Wald CI and the robust likelihood CI. Some results from these studies are presented in this section.

## 5.1 Scenarios in Lin and Wei (1989)

We first evaluated the 12 models examined by Lin and Wei (1989), with the results presented in Table 1. Rows 1-4 of the table are on the omission of relevant covariates from Cox models, rows 5-8 are on the mis-specification of regression forms with possible omission of relevant covariates, and rows 9-12 are on the non-proportional hazard models also with possible omission of relevant covariates. These results indicate that, under these scenarios, the coverage probabilities of both the robust Wald CI and likelihood CI are quite close to the nominal level 0.95. It is interesting to see that their performances are comparable in terms of coverage probabilities and CI width.

*Table 1. Coverage probabilities and width of the robust Wald/likelihood 95% CI for the effect of $Z_1$ under the falsely assumed Cox model $\lambda(t; Z_1, Z_2) = \lambda_0(t)\, exp(\beta_1 Z_1 + \beta_2 Z_2)$.*

|   | True model | Coverage probabilities | | | | CI width | | | |
|---|---|---|---|---|---|---|---|---|---|
|   |   | Robust Wald CI | | Robust Likelihood CI | | Robust Wald CI | | Robust Likelihood CI | |
|   |   | n=100 | n=50 | n=100 | n=50 | n=100 | n=50 | n=100 | n=50 |
| 1 | $\lambda(t) = \exp(.2Z_2 + Z_3)$ | 0.941 | 0.939 | 0.943 | 0.948 | 1.114 | 1.583 | 1.121 | 1.603 |
| 2 | $\lambda(t) = \exp(.2Z_2 + Z_1^2)$ | 0.946 | 0.944 | 0.948 | 0.953 | 1.115 | 1.588 | 1.122 | 1.608 |
| 3 | $\lambda(t) = \exp(Z_1^2)$ | 0.954 | 0.940 | 0.958 | 0.944 | 1.115 | 1.586 | 1.122 | 1.606 |
| 4 | $\lambda(t) = \exp(.2Z_2 + Z_1^2 + Z_3)$ | 0.943 | 0.945 | 0.948 | 0.948 | 1.116 | 1.589 | 1.123 | 1.609 |
| 5 | $\lambda(t) = \exp(1 + .5Z_2)$ | 0.930 | 0.933 | 0.935 | 0.94 | 1.116 | 1.587 | 1.123 | 1.607 |
| 6 | $\lambda(t) = \exp(1 + .5Z_2 + Z_1^2)$ | 0.948 | 0.937 | 0.953 | 0.942 | 1.115 | 1.59 | 1.122 | 1.610 |
| 7 | $\lambda(t) = \log(2 + .5Z_2)$ | 0.958 | 0.955 | 0.959 | 0.960 | 1.114 | 1.584 | 1.121 | 1.603 |
| 8 | $\lambda(t) = \log(2 + .5Z_2 + Z_1^2)$ | 0.948 | 0.949 | 0.950 | 0.950 | 1.115 | 1.583 | 1.122 | 1.603 |
| 9 | $\log T = -.5Z_2 + \phi$ | 0.955 | 0.947 | 0.958 | 0.952 | 1.116 | 1.586 | 1.123 | 1.606 |
| 10 | $\log T = -.5Z_2 - Z_1^2 + \phi$ | 0.948 | 0.950 | 0.948 | 0.955 | 1.114 | 1.59 | 1.121 | 1.610 |
| 11 | $T = \exp(-.5Z_2) + \epsilon$ | 0.942 | 0.941 | 0.944 | 0.943 | 1.115 | 1.584 | 1.122 | 1.604 |
| 12 | $T = \exp(-.5Z_2 - Z_1^2) + \epsilon$ | 0.959 | 0.942 | 0.964 | 0.953 | 1.115 | 1.587 | 1.122 | 1.606 |

NOTE: These models are from Table 1 in Lin and Wei (1989). $Z_1, Z_2$ and $Z_3$ are independent standard normal variables truncated at $\pm 5$ in rows 1-4 and 9-12. $Z_1$ and $Z_2$ are independent standard normal variables truncated at $\pm 1.96$ in rows 5-8. $\phi$ is a zero-mean normal variable with .5 standard deviation. $\epsilon$ is a standard exponential variable. Each entry is based on 1,000 Monte Carlo replications without censoring.

### 5.2 Special case when the events are rare

For illustration purposes, we also consider a clinical trial in which events, such as HIV infections in an HIV prevention trial, are very rare. Here, the only variable of interest is the treatment $Z_1$, which can take values of 0 or 1. The sample size is assumed to be 5000 with a 1:1 randomization ratio, and the trial is event-driven with a total of 20 events. The true model is assumed to be $\lambda(t) = \lambda_0 \exp(\beta_1 Z_1)$ with the baseline event rate $\lambda_0 = 0.004$. The log profile likelihood function can be written as $l(\beta_1) = \beta_1 D - \sum \log\{n_i/m_i + \exp(\beta_1)\}$, where $D$ is the total number of events in the experimental arm, and $(m_i, n_i)$ are the number at risk at the event times for the experimental and control arm, respectively. Since the events are rare (20 events out of 5000 participants), $n_i/m_i \approx 1$ at every event timepoint, therefore the inference of $\beta_1$ primarily depends on the number of events in each arm and not significantly on the timing of these events. This enables the enumeration of all the possible calculated robust likelihood (or Wald) CI by considering the number of events in each arm. Coverage probabilities can then be calculated by summing the probabilities that the calculated CI covers the true parameter value.

The calculated coverage probabilities of the upper 97.5% CI (which corresponds to the two-sided 95% CI) for the robust Wald CI and the robust likelihood CI are shown in Figure 2. The results indicates that the actual coverage probabilities of the robust Wald CI can drop to as low as 0.93, despite the nominal coverage probability being 0.975. In contrast, the coverage probabilities of the robust likelihood CI generally ranges from 0.96 to 0.99, and fluctuates around 0.975 intermittently. Thus, in this specific scenario, the robust likelihood CI demonstrates better performance compared to the robust Wald CI.

*Figure 2. Coverage probabilities of the upper 97.5% CI for the hypothetic HIV prevention trial.*

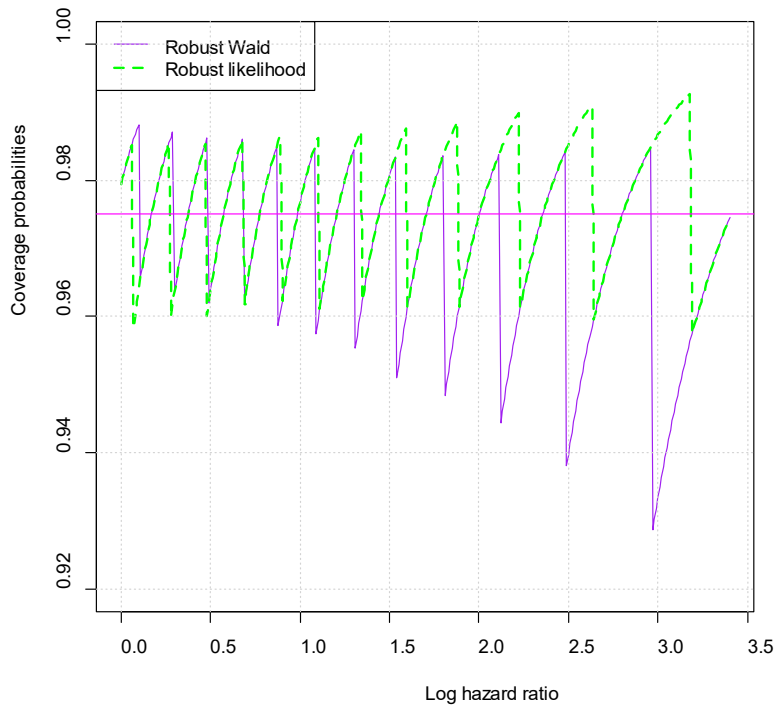

## 6  Example

As we mentioned in the introduction, if we apply the robust Wald CI to the HPTN 084 trial data, the resulting upper CI is not an increasing function of the number of HIV infections in the Cabotegravir arm. We now apply the robust likelihood CI to the HPTN 084 trial, and again vary the number of HIV infections in the Cabotegravir arm, while keeping the number of infections in the TDF-FTC at 36. Since the analysis is at the second interim look, with an O'brien Fleming boundary of 0.00066 (two-sided), we will consider the upper (100-0.033)% CI based on the robust Wald CI and the robust likelihood CI. The results are shown in Figure 3. We can see that the upper CI by the robust Wald CI is not an increasing function of the number of infections, while the robust likelihood CI does not have this issue.

*Figure 3. Hypothetic upper (100-0.033)% CI based on the robust Wald CI and the robust likelihood CI for the HPTN 084 trial, with the number of HIV infections in the Cabotegravir arm varying between 1 and 10, and the number of infections in the TDF-FTC arm fixed as observed.*

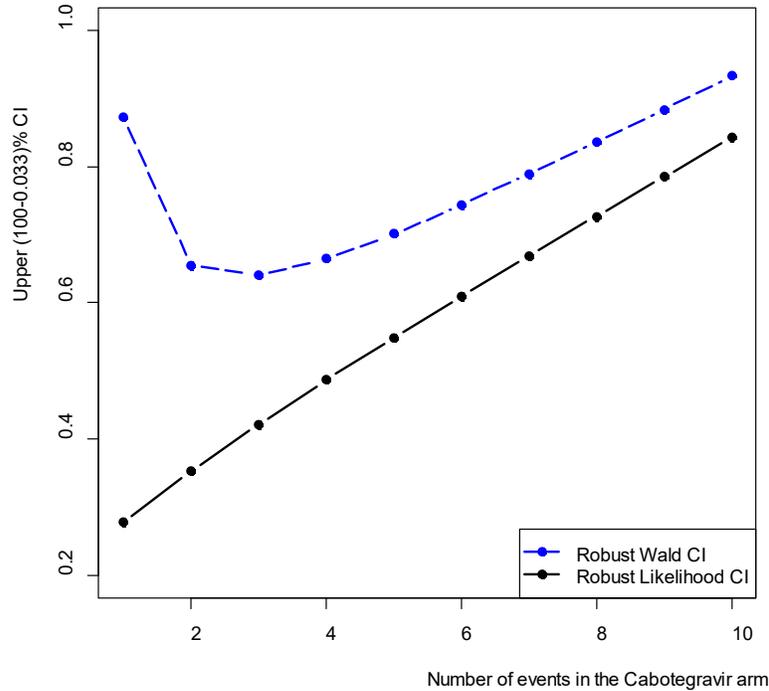

# 7 Discussion

In this article we derived the asymptotic distribution of the usual likelihood ratio test statistic for the misspecified Cox proportional hazard model. When the parameter of interest is one-dimensional, the resulting distribution simplifies a scaled chi-square distribution, where the scaling factor can be consistently estimated. The robust likelihood CI is then obtained by inverting the likelihood ratio test. Using the HPTN 084 trial data, we demonstrated that the proposed robust likelihood CI can have better performance than the robust Wald CI.

In practice, the weighted Cox model is often utilized instead, where the weights may result from survey sampling, where different participants may have different sampling probabilities, or, more commonly, from the propensity score weighting, which has been gaining popularity in observational studies in recent years. Let $v_i$ be the sampling weight of the $i$-th participant. For finite population inferences, we can get the weighted version of $\hat{V}(\beta)$ and $\hat{A}(\beta)$ using the following expressions, similar to what was done in Binder[15] (1992):

$$S^{(r)}(\beta, t) = n^{-1} \sum_{i=1}^{n} v_i Y_i(t) \exp\{\beta^T Z_i(t)\} Z_i(t)^{\otimes r}$$

$$\hat{A}(\beta) = n^{-1} \sum_{i=1}^{n} v_i \delta_i \left\{ \frac{S^{(2)}(\beta, X_i)}{S^{(0)}(\beta, X_i)} - \frac{S^{(1)}(\beta, X_i)^{\otimes 2}}{S^{(0)}(\beta, X_i)} \right\}$$

$$W_i(\beta) = \delta_i \left\{ Z_i(X_i) - \frac{S^{(1)}(\beta, X_i)}{S^{(0)}(\beta, X_i)} \right\} - \sum_{j=1}^{n} \frac{v_j \delta_j Y_i(X_j) \exp\{\beta^T Z_i(X_j)\}}{n S^{(0)}(\beta, X_j)} \times \left\{ Z_i(X_j) - \frac{S^{(1)}(\beta, X_j)}{S^{(0)}(\beta, X_j)} \right\}$$

$$\hat{B}(\beta) = n^{-1} \sum_{i=1}^{n} v_i W_i(\beta)^{\otimes 2}$$

$$\hat{V}(\beta) = \hat{A}^{-1}(\beta) \hat{B}(\beta) \hat{A}^{-1}(\beta)$$

A justification of using the above for the finite population inferences was given by Lin[16] (2000). Then we can use $\sum_{j=1}^{k} \delta_j \chi_j^2(1)$ to approximate the distribution of the likelihood ratio test statistic, where $\delta_j$'s are the eigenvalues of $\hat{V}_{11}(\beta) \left[ \{\hat{A}^{-1}(\beta)\}_{11} \right]^{-1}$.